\documentclass[aps,prd,twocolumn,floatfix,showpacs,superscriptaddress,nofootinbib]{revtex4-2}
\usepackage{amsmath,amsfonts,amssymb,bm}
\usepackage{graphicx}
\usepackage{color}
\usepackage{subfigure}
\usepackage{multirow}
\usepackage{textcomp}
\usepackage{slashed}
\usepackage[bottom]{footmisc}
\definecolor{purple}{rgb}{0.5,0,0.5}
\definecolor{blue}{rgb}{0.0,0,1.0}
\usepackage[colorlinks=true, pdfstartview=FitV, linkcolor=purple, citecolor= purple, urlcolor=blue]{hyperref}
\begin{document}

\title{Quark anomalous magnetic moment and its effects on the $\rho$ meson properties}

\author{Zanbin Xing}
\affiliation{School of Physics, Nankai University, Tianjin 300071, China}

\author{Kh\'epani Raya} 
\affiliation{School of Physics, Nankai University, Tianjin 300071, China}
\affiliation{Instituto de Ciencias Nucleares, Universidad Nacional Aut\'onoma de M\'exico, Apartado Postal 70-543, CdMx 04510, Mexico.}

\author{Lei Chang} 
\affiliation{School of Physics, Nankai University, Tianjin 300071, China}

\date{\today}

\begin{abstract}
A symmetry-preserving treatment of mesons, within a Dyson-Schwinger and Bethe-Salpeter equations approach, demands an interconnection between the kernels of the quark gap equation and meson Bethe-Salpeter equation. Appealing to those symmetries expressed by the vector and axial-vector Ward-Green-Takahashi identitiges (WGTI), we construct a two-body Bethe-Salpeter kernel and study its implications in the vector channel; particularly, we analyze the structure of the quark-photon vertex, which explicitly develops a vector meson pole in the timelike axis and the quark anomlaous magnetic moment term, as well as a variety of $\rho$ meson properties: mass and decay constants, electromagnetic form factors, and valence-quark distribution amplitudes.

\end{abstract}

\maketitle

\section{Introduction}
Quantum chromodynamics (QCD) is regarded as the underlying theory of the nuclear strong interactions, and so hadron physics. Even though its Lagrangian is apparently simple, high-level complexity phenomena take place, such as quark-gluon confinement and the emergence of hadron masses (EHM)~\cite{Roberts:2020hiw}. The Dyson-Schwinger equations (DSE) formalism has proven to be a very robust approach to QCD in the continuum~\cite{Roberts:1994dr,Fischer:2018sdj}, capable of taming its non-perturbative character. Supplemented by the bound-state Bethe-Salpeter (BS) and Faddeev equations~\cite{Faddeev:1961cu,Salpeter:1951sz}, the DSE formalism becomes an ideal platform for the calculation of hadron masses and several structural properties; see, for example Refs.~\cite{Eichmann:2016yit,Cui:2020tdf,Arrington:2021biu,Qin:2019hgk}. The derivation of QCD's DSEs do not require any assumptions on the running coupling, therefore, both perturbative and non-perturbative facets of the strong interactions can be addressed within this formalism. At the same time, the DSE approach is not restricted to a certain domain of current quark masses~\cite{Hilger:2014nma,Raya:2019dnh,Qin:2019hgk,Chen:2020ecu}. Moreover, hadron observables can be traced down to fundamental pieces, namely propagators and vertices, hence maintaining a clear connection to QCD's fundamental degrees of freedom, quarks and gluons~\cite{Raya:2015gva,Chang:2013nia,Chen:2020wuq,Qin:2020rad}. The structure of the DSEs is such that any $n$-point Green function is related to at least one higher order function, therefore yielding an infinite set of coupled integral equations~\cite{Roberts:1994dr}. In order to arrive at a tractable problem, a sensible truncation is needed; thus, systematicity and symmetry principles become imperative~\cite{Binosi:2016rxz,Munczek:1994zz, Bender:1996bb}. Historically, Ward-Green-Takahashi identities (WGTI)~\cite{Ward:1950xp,Green:1953te,Takahashi:1957xn} have been crucial in symmetry-preserving studies of hadron properties. Such relations ensure, among other things, current-conservation and the appearance of Goldstones modes in connection with dynamical chiral symmetry breaking (DCSB)~\cite{Munczek:1994zz, Bender:1996bb}. Furthermore, WGTI impose relationships between propagators and vertices, as well as constraints between the kernels of the one-body and two-body problems~\cite{Chang:2009zb,Qin:2020jig}. Nevertheless, the mathematical form of the one-body and two-body kernels (quark self energy and BS equation kernels), which satisfy such consistency relations, is not always unique~\cite{Chang:2020iut}. Taking advantage of this fact, we
derive a \emph{modified} version of the so called rainbow-ladder (RL) truncation. Subsequently, within a contact interaction (CI) model, we analyze its impact on the structure of the quark-photon vertex (QPV) and $\rho$ meson properties, as derived from their corresponding Bethe-Salpeter equations. More precisely, we shall expose how, in addition to the $\rho$ meson pole in the timelike axis, the structure of the QPV develops an anomalous magnetic moment (AMM) term. It is known that DCSB generates an anomalous chromonagnetic moment for dressed light-quarks, which is large at infrared momenta and generates an electromagnetic moment with commensurate size but opposite sign~\cite{Chang:2010hb,Bashir:2011dp}. Thus, it is a highly desirable feature for the QPV, exhibited quite transparently in our present approach. Its impact on the structural properties of hadrons is put in manifest when studying electromagnetic form factors (EFF): while the elastic EFF of spin-0 mesons is unaltered by the AMM piece, the latter displays from mild to notorious impact on EFF involving spin-1 mesons (and baryons as well)~\cite{Wilson:2011aa,Raya:2018ioy,Raya:2021}.

The manuscript is organized as follows: Section II introduces aspects of constructing symmetry-preserving truncations of quark DSE and meson BS equation, based upon vector and axial-vector WGTIs. In Section III, we present the CI model within the RL truncation and its modified version (MRL). The mass spectrum of light-mesons $\{\pi$, $\rho\}$ is included to compare and contrast. Section IV focuses on the derivation and structure of the QPV. Subsequent sections capitalize on $\rho$ meson structural properties: EFFs of the $\rho$ meson are discussed on Section V, while its valence-quark distribution amplitudes (PDAs) are derived in Section VI. Conclusions and final remarks are presented in Section VII.

\section{Symmetry-preserving truncations}
Within the DSE approach, the properties of valence quark/antiquark bound-states are encoded within solutions of its BS equation. Mesons appear as poles in the corresponding inhomogeneous BS equation~\cite{Qin:2020jig, Chang:2009zb}, which takes the form~\cite{Chang:2009zb,Qin:2020jig}:
\begin{equation}
\label{eq:BSEGen}
    \Gamma_{H}(p;P)=\tilde{\gamma}_H+\int_{q} K^{(2)}(q,p;P)\chi_{H}(q;P)\;,
\end{equation}
where $\chi_{H}(q;P)=S(q_+)\Gamma_{H}(q;P)S(q_-)$ denotes the BS wavefunction, such that: $\Gamma_{H}(q;P)$ corresponds to the BS amplitude of a meson $H$, whose specific structure in terms of Dirac matrices depends on its quantum numbers; $S(q^{\pm})$ represent the quark and antiquark propagators. The two particle irreducible quark/antiquark scattering kernel is denoted by $K^{(2)}(q,p;P)$; and $\tilde{\gamma}_H$, as the BS amplitude, is defined by a specific combination of Dirac matrices that specify the $J^{PC}$ channel ($\tilde{\gamma}_H= \gamma_\mu$ for the vector vertex and $\tilde{\gamma}_H=\gamma_5 \gamma_\mu$ for the axial-vector vertex). Herein, we use the notation $\int_{q}$ to refer to four dimensional Euclidean integral, regularized in a Poincar\'e covariant manner. Finally, the kinematics is defined as follows: $P$ is the total momentum of the quark/antiquark system; $q_+=q+\eta P$ and $q_-=q-(1-\eta)P$, with $\eta\in[0,1]$ defining the relative momentum (in a Poincar\'e covariant framework, no single observable depends on $\eta$).  The fully-dressed quark propagator obeys a DSE of the form:
\begin{equation}
\label{eq:quarkPropGen}
    S^{-1}(p)=[S^{(0)}(p)]^{-1}+\int_{q} K^{(1)}(q,p)S(q)\;,
\end{equation}
where $S^{(0)}(p)=[i \gamma \cdot p+m^{\text{bm}}]^{-1}$ corresponds to the bare quark propagator, with a Lagrangian current quark mass $m^{\text{bm}}$, and $K^{(1)}(q,p)$ the one-body kernel. The above equation is often referred to as gap equation. In both Eqs.~(\ref{eq:BSEGen}, \ref{eq:quarkPropGen}), and throughout the rest of the manuscript, we supress all remormalization constants, as well as color and flavor indices, for notational convenience.

As has been pointed out in the Introduction, a practical way to construct and relate $K^{(1)}$ and $K^{(2)}$, relies upon WGTIs. The chiral limit vector and axial vector WGTIs take the form:
\begin{eqnarray}\label{eq:vWGTI}
i P_{\mu}\Gamma_{\mu}(k;P)&=&S^{-1}(k_{+})-S^{-1}(k_{-})\;.\\
\label{eq:avWGTI}
P_{\mu}\Gamma_{5\mu}(k;P)&=&S^{-1}(k_{+})i\gamma_{5}+i\gamma_{5}S^{-1}(k_{-})\;,
\end{eqnarray}
where $\Gamma_{\mu}$ and $\Gamma_{5\mu}$ are the vector and axial-vector vertices. Now consider the gap equation in QCD~\cite{Roberts:1994dr}:
\begin{equation}
\label{eq:quarkPropQCD}
    S^{-1}(p)=[S^{(0)}(p)]^{-1}+\frac{4}{3}g^2 \int_{q} D_{\mu\nu}(p-q) \gamma_\mu S(q) \Gamma_\nu(p,q)\;,
\end{equation}
where $g$ is the Lagrangian coupling constant; $D_{\mu\nu}$ and $\Gamma_\nu$ are the fully-dressed gluon propagator and quark-gluon vertex (QGV), respectively. Thus we can identify
\begin{equation}
\label{eq:K1}
    K^{(1)}(q,p)=\frac{4}{3} g^2 D_{\mu\nu}(p-q)\gamma_\mu \otimes \Gamma_\nu(q,p)\;.
\end{equation}
If we restrain ourselves to the tree-level QGV $\Gamma_\nu \to \gamma_\nu$, hence neglecting all the rich structure that the fully-dressed vertex might have~\cite{Albino:2018ncl,Sultan:2018qpx,Bashir:2011dp}, and replace the gluon propagator by an effective one, $D_{\mu\nu} \to D_{\mu\nu}^{\text{eff}}$, to compensate the missing pieces in $\Gamma_\nu$~\cite{Qin:2011dd,Maris:1999nt}, the one-body kernel becomes
\begin{equation}
\label{eq:K1RL}
    K^{(1)}(q,p)=\frac{4}{3}g^2 D_{\mu\nu}^{\text{eff}}(p-q)\gamma_\mu \otimes \gamma_\nu\;.
\end{equation}
The two-body kernel $K^{(2)}$ can be obtained by combining Eqs.~(\ref{eq:BSEGen}-\ref{eq:avWGTI}) and Eq.~\eqref{eq:K1RL}, which yields
\begin{eqnarray}
\label{eq:vecIdWG}
\int_{q}K^{(2)}(q,p;P)[S(q_+)-S(q_-)]= \hspace{2cm}&\\ \nonumber
-g^2\frac{4}{3}\int_{q} D_{\mu\nu}^{\text{eff}}(p-q)\gamma_\mu[S(q_+)-S(q_-)]\gamma_\nu \;,& \\
\label{eq:axIdWG}
\int_{q}K^{(2)}(q,p;P)[S(q_+)\gamma_5+\gamma_5S(q_-)]=\hspace{2cm}& \\ \nonumber
-g^2\frac{4}{3}\int_{q} D_{\mu\nu}^{\text{eff}}(p-q)\gamma_\mu[S(q_+)\gamma_5+\gamma_5 S(q_-)]\gamma_\nu \;.&
\end{eqnarray}
Then the simplest choice that satisfies the vector and axial-vector WGTIs symmetry constraints is
\begin{equation}
\label{eq:defRL}
    K^{(2)}(q,p;P)=-\frac{4}{3}g^2 D_{\mu\nu}^{\text{eff}}(p-q)\gamma_\mu \otimes \gamma_\nu=-K^{(1)}(q,p;P)\;.
\end{equation}
Such kernels define the RL truncation: $K^{(1)}$ refers to the `\emph{rainbow}' part, while $K^{(2)}$ corresponds to the `\emph{ladder}' piece. Notably, this simple choice is sufficient to ensure the appearance of pions as Goldstone bosons of DCSB~\cite{Munczek:1994zz,Bender:1996bb}, while also being a sensible approximation to compute its structural properties~\cite{Ding:2019lwe, Raya:2015gva, Chang:2013nia}. 

It is worth pointing out that the solution to Eqs.~(\ref{eq:vecIdWG}-\ref{eq:axIdWG}) is not unique. In fact, given the rainbow approximation for the one body problem, it is possible to derive a fully consistent symmetry-preserving two body kernel that extends beyond the RL truncation, as it is discussed in Ref.~\cite{Chang:2020iut,Ding:2018xwy}. In addition to the ladder part, other terms of the form
\begin{equation}
    K^{(2)}(q,p;P)=\tilde{D}^j(q,p) \tilde{\Gamma}_{j}(q,p)\otimes \tilde{\Gamma}_{j}(q,p) F^j(q,p)\;,
\end{equation}
can be added to the two-body kernel~\cite{Dai:1990ap}. Herein, $F^j$ are Lorentz invariant scalar functions, $\tilde{\Gamma}_{j}$ are different combinations of Dirac matrices, and $\tilde{D}^j$ are tensor structures that might be needed to contract Lorentz indices. The particular choices $\tilde{\Gamma}_{j}(q,p) = \{ \mathbb{I},\;\gamma_5,\;i/\sqrt{6}\sigma_{\mu\nu}\}$ and $\tilde{D}^j(p,q)F^j(p,q)  \to \xi \tilde{D}$, the latter being reduced to a simple constant, preserve the consistency constraints and so Eqs.~(\ref{eq:vecIdWG}-\ref{eq:axIdWG}). Thus, the inhomogeneous BS equation becomes
\begin{eqnarray}\nonumber
    \Gamma_{H}(q;P)&=&\tilde{\gamma}_H-\frac{4}{3}g^2\int_{q} D_{\mu\nu}^{\text{eff}}(p-q)\gamma_\mu \chi_H(q;P) \gamma_\nu\\
    &+& \xi \tilde{D} \int_{q}  \tilde{\Gamma}_{j} \chi_H(q;P)\tilde{\Gamma}_{j}\;.
\label{eq:inBSEERL}
\end{eqnarray}
The product $\xi \tilde{D}$ sets the strength of the non-ladder (NL) term. In the next section we shall discuss about the truncation herein derived, Eq.~\eqref{eq:inBSEERL}, within a vector-vector symmetry-preserving contact interaction (CI) model of QCD~\cite{GutierrezGuerrero:2010md, Roberts:2010rn,Wilson:2011aa}.

 \medskip
\section{Contact Interaction model}
Let us recall the quark gap equation in the RL truncation:
\begin{eqnarray}
\label{eq:quarkGapRL}
    S^{-1}(p)=[S^{(0)}(p)]^{-1}+\frac{4}{3}\int_{q} g^2 D_{\mu\nu}^{\text{eff}}(p-q) \gamma_\mu S(q) \gamma_\nu\;.\hspace{0.3cm}
\end{eqnarray}
Clearly, the quark DSE decouples from the QGV and gluon DSEs. The only remaining ingredient is $D_{\mu\nu}^{\text{eff}}(p-q)$, the effective gluon propagator. This piece is supposed to compensate for all the missing pieces in the QGV~\cite{Maris:1999nt,Qin:2011dd}, often requiring an artificial enhancement in the infrared~\cite{Sultan:2018qpx}. Thus we appeal to the illustrative CI model introduced in Refs.~\cite{GutierrezGuerrero:2010md, Roberts:2010rn},
\begin{equation}
    \label{eq:gluonprop}
    g^2D_{\mu\nu}^{\text{eff}}(p-q) \to \frac{1}{m_G^2}\delta_{\mu\nu}\;,
\end{equation}
where $m_G = 0.132$ GeV is an infrared mass scale. Besides preserving the relevant symmetries, the CI model typically yields semialgebraic expressions and captures the non-perturbative traits of QCD~\cite{GutierrezGuerrero:2010md, Roberts:2010rn,Wilson:2011aa}. In addition, the CI model produces sensible results for the hadron mass spectrum~\cite{Yin:2021uom, Gutierrez-Guerrero:2019uwa,Yin:2019bxe,Chen:2012qr}, including tetraquarks~\cite{Bedolla:2019zwg}, while also providing crucial benchmarks for many hadron structural properties~\cite{Wilson:2011aa,Raya:2018ioy,Raya:2017ggu,Bedolla:2016yxq, Segovia:2015hra, Segovia:2014aza}. The following sections are dedicated to illustrate some of the implications of the gluon model Ansatz from Eq.~\eqref{eq:gluonprop}, in the RL and MRL truncations.

\subsection{Contact Interaction in RL truncation}
In the CI model, the DSE for the quark propagator adopts the form
\begin{equation}
\label{eq:quarkPropCI}
    S^{-1}(p)=[S^{(0)}(p)]^{-1}+\frac{4}{3 m_G^2} \int_{q}  \gamma_\mu S(q) \gamma_\mu\;,
\end{equation}
while the homogeneous meson BS equation is written as
\begin{equation}
\label{eq:mesonBSECI}
    \Gamma_H(p;P)=-\frac{4}{3 m_G^2} \int_{q}  \gamma_\mu \chi_H(q;P) \gamma_\mu\;.
\end{equation}
A general solution of Eq.~\eqref{eq:quarkPropCI} is $S^{-1}(p)=i\gamma\cdot p + M$, which exhibits a momentum independent mass function, $M$. The gap equation becomes
\begin{equation}
M=m+\frac{M}{3\pi^2 m_{G}^{2}}\int_{0}^{\infty}ds\frac{s}{s+M}\;,
\end{equation}
which requires a regularization procedure. Following standard literature,~\cite{GutierrezGuerrero:2010md}, we perform a proper time regularization~\footnote{A substraction scheme is also possible, as illustrated in Ref.~\cite{Serna:2017nlr}.}:
\begin{equation}
\frac{1}{s+M^2}=\int_{0}^{\infty}d\tau e^{-\tau (s+M^2)}\rightarrow\int_{\tau_{\text{UV}}^2}^{\tau_{\text{IR}}^2}d\tau e^{-\tau (s+M^2)}.
\end{equation}
The mass scale $\Lambda_{\text{IR}}:=1/\tau_{\text{IR}} = 0.24$ GeV guarantees confinement by ensuring the absence of quark production thresholds, while $\Lambda_{\text{UV}}:=1/\tau_{\text{UV}}=0.905$ GeV represents an ultraviolet cut-off, setting the scale of all dimensioned quantities because the theory is non renormalizable. Therefore, the mass function can be obtained by solving
\begin{eqnarray}
M&=&m+\frac{M}{3\pi^2 m_{G}^{2}}\mathit{C}^{iu}(M^2)\;,\\
\frac{\mathit{C}^{iu}(M^2)}{M^2}&=&\Gamma(-1,M^2\tau_{\text{UV}}^2)-\Gamma(-1,M^2\tau_{\text{IR}}^2)\;,
\end{eqnarray}
where $\Gamma(a,z)$ is the incomplete Gamma function.

Concerning the meson BS equation in the CI, Eq.~\eqref{eq:mesonBSECI},  it is clear that a dependence on the relative momentum is forbidden by the interaction. Then, the pseudoscalar and vector meson BS amplitudes adopt the form
\begin{eqnarray}
\label{eq:BSApi}
\Gamma_{0^-}(P) &=& \gamma_5 \left[ i E_{0^-}(P) + \frac{\gamma \cdot P}{M} F_{0^-}(P) \right]\;,\\
\label{eq:BSArho}
\Gamma^{1^-}_{\mu}(P)&=&\gamma_{\mu}^{T}E_{1^-}(P)+\frac{1}{M}\sigma_{\mu\nu}P_{\nu}F_{1^-}(P)\;,
\end{eqnarray}
where $\gamma_{\mu}^{T}=\gamma_{\mu}-\frac{ \gamma \cdot P \;
 }{P^{2}} \, P_{\mu}$. In a RL treatment of the CI, $F_{1^-}(P) = 0$ (an analogous result holds for the axial-vector meson). Appealing to the WGTIs from Eqs.~(\ref{eq:vWGTI}, \ref{eq:avWGTI}), and contracting with $P_\mu$, one arrives at the chiral limit identities ($P^2=0=-m_H^2$):
\begin{eqnarray}\nonumber
M&=&\frac{8M}{3m_{G}^{2}}\int_{q}\left[\frac{1}{q^2+M^2}+\frac{1}{(q-P)^2+M^2}\right]\;,\\
0&=&\int_{q}\left[\frac{P\cdot q}{q^2+M^2}-\frac{P\cdot(q-P)}{(q-P)^2+M^2}\right],
\end{eqnarray}
which must be satisfied even after regularization, therefore imposing
\begin{eqnarray}
\label{eq:CIvecID1}
M&=&\frac{16M}{3m_{G}^{2}}\int_{q}\frac{1}{q^2+M^2}\\
\label{eq:CIaxID1}
0&=&\int_{q}\frac{\frac{q^2}{2}+M^2}{(q^2+M^2)^2}.
\end{eqnarray}
Notice that Eq.~\eqref{eq:CIvecID1} is merely the chiral-limit gap equation, whereas Eq.~\eqref{eq:CIaxID1} entails that the axial-vector WGTI is satisfied if, and only if, the model is regularized so as to ensure there are no quadratic or logarithmic divergences~\cite{Roberts:2010rn}. 

\subsection{Contact Interaction in Modified RL}
Let us now consider the BS equation in the MRL truncation, Eq.~\eqref{eq:inBSEERL}. Supplemented by the effective gluon in the CI model, Eq.~\eqref{eq:gluonprop}, and setting $\tilde{D}=4/3m_G^2$, the modified homogeneous BS equation becomes
\begin{eqnarray}
\label{eq:CIinMRL}
\Gamma_H(P)=-\frac{4}{3m_{G}^{2}}\int_{q}\left[\gamma_{\mu}\chi_H(P)\gamma_{\mu} - \xi \tilde{\Gamma}_{j}\chi_H(P)\tilde{\Gamma}_{j}\right],\hspace{0.6cm} 
\end{eqnarray}
where we have made evident the momentum-independent nature of the BS amplitude and quark mass function. Recalling that $\tilde{\Gamma}_{j} = \{ \mathbb{I},\;\gamma_5,\;i/\sqrt{6}\sigma_{\mu\nu}\}$ and performing Fierz transformatioon, it can be shown that the NL term, the one proportional to $\xi$, can be rewritten as $\frac{1}{3}\sigma_{\alpha\beta}\text{tr}_{D}[\sigma_{\alpha\beta}\chi_{H}(P)]$ and does not contribute in the case of pseudoscalar mesons; it impacts, however, the vector meson case. Particularly, the $F_{1^-}(P)$ BS amplitude in Eq.~\eqref{eq:BSArho} is no longer zero. This is a crucial difference with respect to the well known CI-RL truncation, which is clearly recovered in the limit case $\xi = 0$.

The BS equation can be recast into an eigenvalue equation by finding proper projectors that decouple $E_{1^-}$ and $F_{1^-}$ in Eq.~\eqref{eq:BSArho}, such that Eq.~\eqref{eq:CIinMRL} yields
\begin{equation}\label{kerneln}
\begin{aligned}
\left[\begin{array}{c}E_{1^{-}}(P)\\F_{1^{-}}(P)\end{array}\right]=\frac{1}{3\pi^2m_G^2}\left[\begin{array}{cc}K_{EE}^{1^{-}}&K_{EF}^{1^{-}}\\K_{FE}^{1^{-}}&K_{FF}^{1^{-}}\end{array}\right]\left[\begin{array}{c}E_{1^{-}}(P)\\F_{1^{-}}(P)\end{array}\right],
\end{aligned}
\end{equation}
where the integration kernels, $K_{ij}:=3\pi^2m_G^2 \mathcal{K}_{ij}$, are written as
\begin{eqnarray}
\mathcal{K}_{EE}^{1^-}=-P^{2}\tilde{I}(P^2)\;&,&\;\mathcal{K}_{EF}^{1^-}=-\frac{P^2}{2}I(P^2)\;,\\
\mathcal{K}_{FE}^{1^-}=\eta_\xi I(P^2)\;&,&\;\mathcal{K}_{FF}^{1^-}=\eta_\xi \left(I(P^2)-\frac{P^2}{M^2}\tilde{I}(P^2) \right),\hspace{0.5cm}
\end{eqnarray}
 the integrals:
\begin{eqnarray}
&&I(Q^2):=\frac{1}{3\pi^2m_G^2}\int_{0}^{1}d\alpha\bar{\mathcal{C}}_{1}^{\text{iu}}(\omega(M^2,\alpha,Q^2))\;,\\
&&\tilde{I}(Q^2):=\frac{1}{3\pi^2m_G^2}\int_{0}^{1}d\alpha\alpha(1-\alpha)\bar{\mathcal{C}}_{1}^{\text{iu}}(\omega(M^2,\alpha,Q^2)\;,\hspace{0.5cm}
\end{eqnarray}
the argument $\omega(M^{2},\alpha,P^{2}):=M^2+\alpha(1-\alpha)P^2$ and, finally,
\begin{eqnarray}
\nonumber
&&\eta_\xi := \frac{2M^2\xi}{3}\;,\;\bar{\mathcal{C}}_{1}^{\text{iu}}(\omega):=-\frac{d}{d\omega}\mathcal{C}^{\text{iu}}(\omega)\;.
\end{eqnarray}
Physical solutions of Eq.~\eqref{kerneln} are only valid for discrete values of $P^2 = -m_H^2$. The smallest $P^2$ that satisfies the eivengalue equation yields the ground-state meson mass. To produce other physical observables, the BS amplitudes must be canonically normalized, according to the condition
\begin{eqnarray}
1=2N_{c}\frac{d}{dP^{2}}\frac{1}{3}\text{tr}_{D}\int_{q}\Gamma_{\mu}(-K)S(q)\Gamma_{\mu}(K)S(q-P)|_{K=P}\;.\hspace{0.5cm}
\end{eqnarray}
Computing the vector meson decay constant, and its tensor counter part, is straighforward from the expressions
\begin{eqnarray}
f_{v}m_{v}&=&\frac{N_{c}}{3}\int_{q}\text{tr}_{D}\gamma_{\mu}S(q)\Gamma_{\mu}S(q-P)\;,\\
f_v^\perp m_v^2&=& \frac{N_{c}}{3} \int_q \sigma_{\mu\nu} P_\nu S(q)\Gamma_{\mu}S(q-P)\;.
\end{eqnarray}
The masses and decay constants of $\pi$ and $\rho$ mesons are collected in Table~\ref{tab:masscball}, in both RL and MRL truncations. The parameter $\xi=0.6$ has been tuned to produce $m_\rho = 0.770$ GeV, as compared with $m_\rho = 0.929$ GeV, obtained in the CI-RL; on the other hand, the leptonic decay constant is quite similar in both truncations, practically independent of $\xi$. Canonically normalized BS amplitudes the $\rho$ meson are listed in Table~\ref{tab:masscball} as well. As already suggested, the static properties of pseudoscalar and axial vector mesons remain unchanged, since the NL term in the BS kernel does not contribute in such channels; we follow standard literature, \emph{e.g.}~\cite{Roberts:2010rn,Roberts:2011wy}, to compute the static properties of pion.
\begin{table}[ht!]
\caption{\label{tab:masscball} CI model results in the RL and MRL truncations. The model parameters: $m_{G}=0.132\;\text{GeV}$, $\Lambda_{\text{UV}}=0.905\;\text{GeV}$, $\Lambda_{\text{IR}}=0.240\;\text{GeV}$ and $\xi=0.6$. The current quark mass is set to $m=m_{u/d} = 7$ MeV, which yields $M = 0.368$ GeV. }
\begin{tabular}{c|cc|ccc|cc}
\hline
 &$m_{\pi}$  & $f_{\pi}$   &$m_{\rho}$ &$f_{\rho}$ & $f_{\rho}^\perp$ & $E_\rho$ & $F_\rho$\\
\hline
CI-RL  & 0.140 & 0.101&  0.929& 0.129 & 0.133 & 1.531 & -\\
CI-MRL & 0.140 & 0.101&  0.770& 0.125 & 0.134 & 1.230 & 0.503 \\
\hline
\end{tabular}
\end{table}

\section{Quark-Photon vertex in Modified RL}
The quark-photon vertex in the CI-MRL satisfies  the inhomogeneous BS equation
\begin{eqnarray}\label{qfv}
\Gamma_\mu^\gamma(Q)=\gamma_\mu&&-\frac{4}{3m_G^2}\int_q \gamma_\alpha S(q)\Gamma_\mu(Q)S(q-Q)\gamma_\alpha\nonumber\\
&&+\frac{4\xi}{3m_G^2}\int_q \tilde{\Gamma}_j S(q)\Gamma_\mu(Q)S(q-Q)\tilde{\Gamma}_j
\end{eqnarray}
A general solution of Eq.~\eqref{qfv} admits a decomposition in terms of 3 tensor structures, namely:
\begin{eqnarray}
\Gamma_\mu^\gamma(Q)=V_1(Q^2)\gamma_\mu^L+V_2(Q^2)\gamma_\mu^T+V_3(Q^2)\gamma_\mu^A\;,&&\nonumber \\
\gamma_\mu^L=\gamma_\mu-\gamma_\mu^T\;,\gamma_\mu^T=\gamma_\mu-\frac{\slashed{Q}Q_\mu}{Q^2}\;,\gamma_\mu^A=\frac{\sigma_{\mu\nu}Q_\nu}{M}\;.&&\label{eq:qfvstructure}
\end{eqnarray}
This simplicity is due to the momentum independent nature of the CI model. By solving Eq.~$(\ref{qfv})$, plainly one obtains that the longitudinal piece is simply
\begin{equation}
 V_1(Q^2):=P_L(Q^2)=1\;,   
\end{equation}
while the transverse dressing functions are expressed as
\begin{eqnarray}
&&V_2(Q^2)=\left[P_T^{-1}(Q^2)+\eta_\xi\frac{Q^2}{2}\frac{I^2(Q^2)}{1-\eta_\xi \bar{I}(Q^2)}\right]^{-1}\;,\\
&& \label{eq:qfv38} V_3(Q^2)= \frac{\eta_\xi I(Q^2)}{P_T^{-1}(Q^2)\left[1-\eta_\xi \bar{I}(Q^2)\right]+\eta_\xi \frac{Q^2}{2} I^2(Q^2) }\;,\hspace{0.5cm} 
\end{eqnarray}
\text where $P_T(Q^2):=[1+Q^2\tilde{I}(Q^2)]^{-1}$, and
\begin{eqnarray}
\bar{I}(Q^2)&&:=I(Q^2)-\frac{Q^2}{M^2}\tilde{I}(Q^2) \;.
\end{eqnarray}
The  $V_2(Q^2)$ dressing function produces a timelike vector meson pole at $Q^2 = -m_\rho^2$, while $V_3(Q^2)$ can be regarded as a profile function for the anomalous magnetic moment term (AMM), $\gamma_\mu^A$.  Notably, in the $Q^2 \to 0$ limit, the fully dressed QPV becomes
\begin{equation}
    \Gamma_\mu^\gamma(Q) \overset{Q^2\to 0}{=} \gamma_\mu + \frac{\sigma_{\mu\nu}Q_\nu}{2M} \left(\frac{2\eta_\xi I(0)}{1-\eta_\xi I(0)} \right)\;,
\end{equation}
such that, with the parameters listed in Table~\ref{tab:masscball}, one gets
\begin{equation}
\label{eq:zetaMRL}
    \zeta_{\text{MRL}}:=\left(\frac{2\eta_\xi I(0)}{1-\eta_\xi I(0)} \right) \approx 0.19.
\end{equation}

\begin{figure}[t!]
\includegraphics[width=8.6cm]{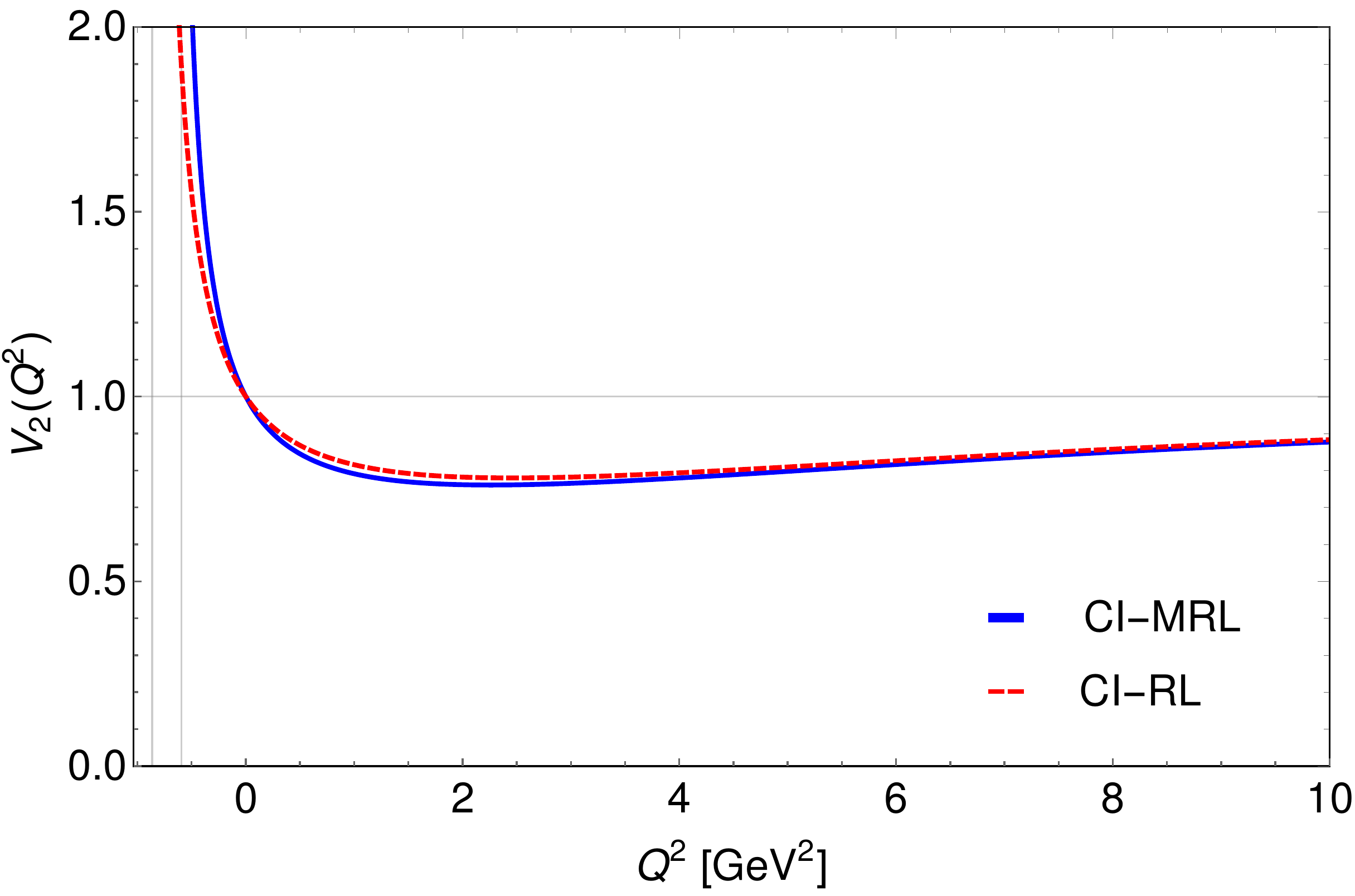}\\
\includegraphics[width=8.6cm]{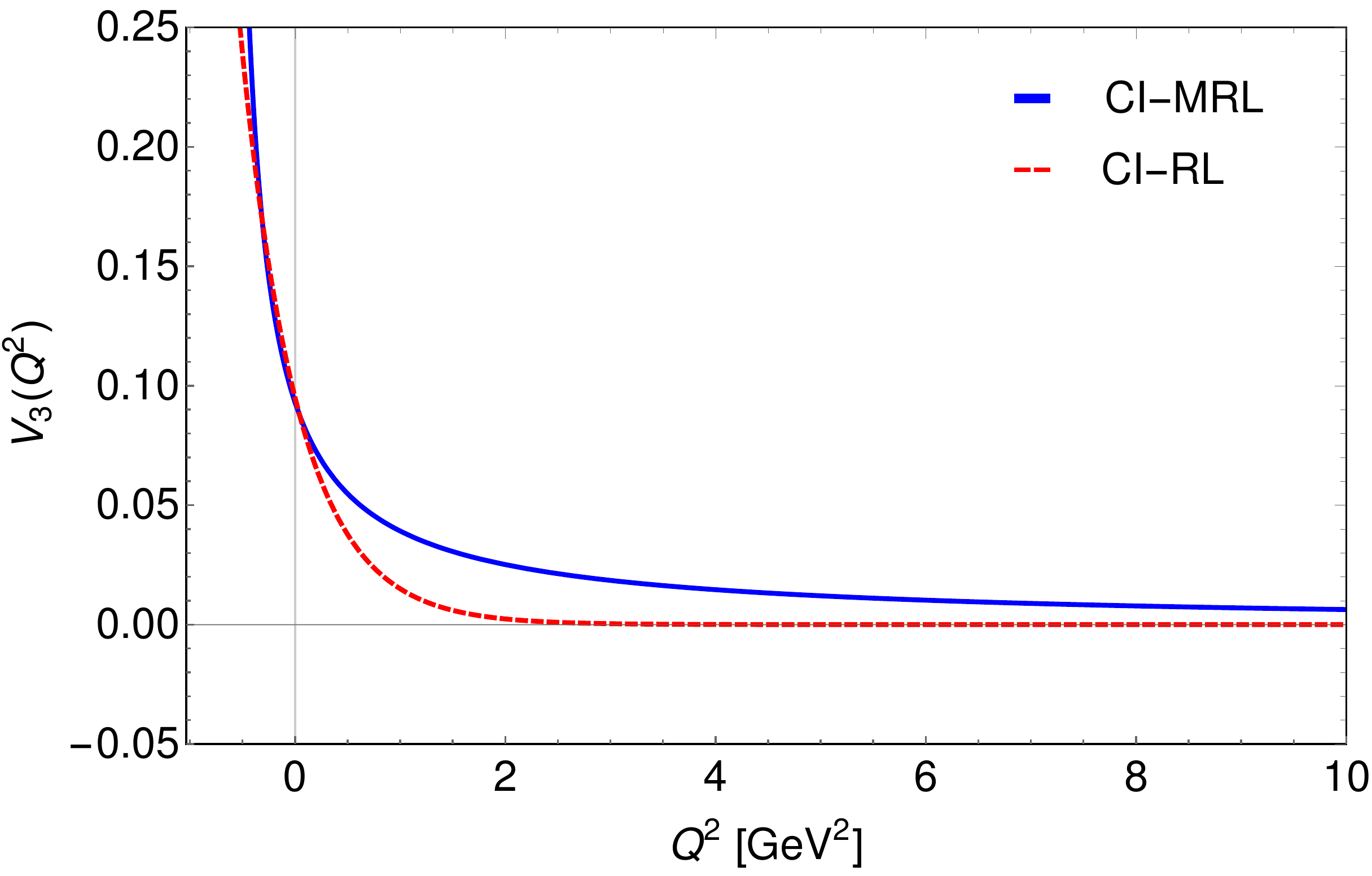}
\caption{\textit{Quark-photon vertex dressing functions} as obtained from Eqs.~(\ref{eq:qfvstructure}-\ref{eq:qfv38}). The upper panel displays the dressing function $V_2$, associated with the $\rho$ meson pole. Vertical grid lines indicate the location of the poles, $Q^2 = -m_\rho^2$, such that $m_\rho = 0.929$ GeV in CI-RL and $m_\rho = 0.770$ GeV in CI-MRL. Notably,  $V_2(Q^2)=P_T(Q^2)$ in the CI-RL case. The lower panel depicts the dressing of the AMM piece of the QPV. Strictly speaking, the CI-RL case implies $V_3(Q^2) = 0$; therefore, the Gaussian Ansatz from Eq.~\eqref{eq:AMMRL} and Refs.~\cite{Wilson:2011aa,Raya:2018ioy,Raya:2021} is displayed instead ($\zeta_{\text{CI-RL}}=\zeta_{\text{CI-MRL}}=0.19$). \label{fig:vertexdress}}
\end{figure}

The automatic incorporation of the AMM piece to the QPV, via inhomogeneous BS equation, is a desirable feature of the MRL truncation. The CI model plainly exposes it. This characteristic is not present in the RL approximation~\cite{Chang:2010hb}, and so in the CI-RL case; the latter corresponding to the case $\xi = 0$, which implies $V_2(Q^2) = P_T(Q^2)$ and $V_3(Q^2) = 0$~\cite{Roberts:2010rn, Roberts:2011wy}. Within the CI-RL approach, the AMM term is typically added by hand, such that the behavior of $V_3(Q^2)$ is modeled according to the exponential Ansatz~\cite{Wilson:2011aa,Raya:2018ioy,Raya:2021}:
\begin{equation}
\label{eq:AMMRL}
    V_3^\text{CI-RL}(Q^2) :=\frac{\zeta_{\text{CI-RL}}}{2}\text{exp}[-Q^2/(4 M^2)]\;,
\end{equation}
where $\zeta_{\text{RL}} \in [0,\;0.5]$ is a strength parameter, fully compatible with $\zeta_{\text{MRL}} \approx 0.19$. Thus, when dealing with the CI-RL truncation, we modify the QPV to account for the AMM, such that
\begin{equation}
\label{eq:EnhanceQPV}
    \Gamma_\mu^{\gamma} \overset{\text{RL}}{:=} \gamma_\mu^L + \gamma_\mu^T P_T(Q^2) + \gamma_A  V_3^\text{CI-RL}(Q^2)\;.
\end{equation}
The vertex dressing $V_2(Q^2)$ is compared with its CI-RL counterpart, $P_T(Q^2)$, in the upper panel of Fig.~\ref{fig:vertexdress}. The presence of the vector meson pole at $Q^2 = -m_\rho^2$ is a natural artifact of the RL truncation when obtaining the QPV through its corresponding BS equation~\cite{Maris:2002mz,Eichmann:2019bqf,MiramontesLopez:2021oul,Xu:2021mju}. It is then expected that electromagnetic form factors, in the vicinity of $Q^2 = 0$, be affected by the $\rho$ pole (for instance, it influences the associated charge radius)~\cite{Maris:1999bh}; but the effects should be otherwise immaterial at large space-like momenta~\cite{Maris:1999bh,Raya:2015gva}. The $Q^2$ profiles of the AMM dressing functions, in both truncations, are displayed in the lower panel of Fig.~\ref{fig:vertexdress}. It is clear that $V_3(Q^2)$ enhances the strength of the QPV in the low $Q^2$ domain, and its contribution vanishes as $Q^2$ grows; the CI-MRL case is power-law suppresed, in contrast with the Gaussian Ansatz in Eq.~\eqref{eq:AMMRL}. In any case, the enhancement-damping patterns manifest in electromagnetic form factors involving spin-1 mesons, but are strictly ruled out in elastic form factors of spin-0 mesons~\cite{Wilson:2011aa,Raya:2018ioy,Raya:2021}.

Finally, it is worth noticing that the QPV defined through Eqs.~(\ref{eq:qfvstructure}, \ref{eq:qfv38}) exhibits the correct asymptotic limit, \emph{i.e.} $\Gamma_\mu^\gamma (Q) \to \gamma_\mu$ as $Q^2 \to \infty$; therefore exposing that a dressed quark becomes pointlike for a large-$Q^2$ probe~\cite{Bermudez:2017bpx}. In the next section, we study the $\rho$ meson elastic EFF, capitalizing on how it is affected by the anomalous magnetic moment.

\medskip
\section{Electromagnetic form factors}
The coupling of a photon to a $1^-$ meson is characterized by three elastic form factors, such that the $\gamma\rho$ vertex can be expressed as~\cite{Roberts:2011wy}:
\begin{eqnarray}
\label{eq:lambdavertex1}
\Lambda_{\lambda,\mu\nu}(K,Q)&=&\sum_{j=1}^3 T^{(j)}_{\lambda,\mu\nu}(K,Q) F_j(Q^2) \;\\\label{eq:lambdavertex2}
T^{(1)}_{\lambda,\mu\nu}(K,Q)&=&2K_\lambda P_{\mu\alpha}^T(p_i)P_{\alpha\nu}^T(p_f)\;,\\ \nonumber
T^{(2)}_{\lambda,\mu\nu}(K,Q)&=&\left[Q_\mu-p_{\mu}^i\frac{Q^2}{2m_\rho^2}\right]P_{\lambda\nu}^T(p_f) \nonumber \\
&-&\left[Q_\nu+p_{\nu}^f\frac{Q^2}{2m_\rho^2}\right]P_{\lambda\mu}^T(p_i)\;,\\ \nonumber
T^{(3)}_{\lambda,\mu\nu}(K,Q)&=&\frac{K_\lambda}{m_\rho^2}\left[Q_\mu-p_{\mu}^i\frac{Q^2}{2m_\rho^2}\right]\left[Q_\nu+p_{\nu}^f\frac{Q^2}{2m_\rho^2}\right]\;.
\end{eqnarray}
The kinematic variables are defined as follows: $p^i = K-Q/2$  and $p^f = K+Q/2$ denote the incoming and outgoing meson momenta, respectively, and $Q$ is the photon momentum; the on-shell conditions, $p_i^2=p_f^2=-m_\rho^2$, impose $K\cdot Q=0$, $K^2=-m_\rho^2-Q^2/4$; and, as explained elsewhere~\cite{Roberts:2011wy}, a symmetry-preserving treatment demands the WGTIs:
\begin{eqnarray*}
p_\mu^i \Lambda_{\lambda,\mu\nu} = p_\nu^f \Lambda_{\lambda,\mu\nu} = Q_\lambda \Lambda_{\lambda,\mu\nu}=0\;.
\end{eqnarray*}
In the impulse approximation~\cite{Roberts:1994hh}, meson EFFs are completely described in terms of quark propagators, BS amplitudes and the QPV. Particularly, the elastic form factor of the $\rho$ meson reads
\begin{eqnarray}
\label{eq:impulse}
\Lambda_{\lambda,\mu\nu}(Q^2)&&=2 N_c\ \text{tr}\int_q \left[\Gamma_\nu^\rho(-p_f)S(q+p_f)\right.\nonumber\\ 
&&\left.\times i\Gamma_\lambda^\gamma(Q)S(q+p_i)\Gamma_\mu^\rho(p_i)S(q)\right]\;,
\end{eqnarray} 
where $\Gamma_\mu^\rho$ is the BS amplitude and $\Gamma_\lambda^\gamma(Q)$ is the QPV. Form factors are obtained from Eqs.~(\ref{eq:lambdavertex1}, \ref{eq:lambdavertex2}, \ref{eq:impulse}), choosing appropriate projectors that decouple them from $\Lambda_{\lambda,\mu\nu}$. The particular expressions for $F_i(Q^2)$ are shown explicitly in Appendix A. It turns out convenient to relate the form factors $F_j(Q^2)$ with the electric, magnetic and quadrupole form factors, respectively
\begin{eqnarray}
&&G_E(Q^2)=F_1(Q^2)+\frac{2}{3}\frac{Q^2}{4M_\rho^2}G_Q(Q^2)\;,\\
\nonumber &&G_M(Q^2)=-F_2(Q^2)\;,\\
\nonumber&&G_Q(Q^2)=F_1(Q^2)+F_2(Q^2)+F_3(Q^2)\left[1+\frac{Q^2}{4M_\rho^2}\right]\;.
\end{eqnarray}
Naturally, the $Q^2 \to 0$ limit defines the charge, magnetic and quadrupole moments:
\begin{eqnarray}
&&G_E(Q^2\to 0)=1\;,\\
&&G_M(Q^2\to 0)=\mu_\rho \;,\nonumber \\
&&G_Q(Q^2\to 0)=\mathcal{Q}_\rho\;. \nonumber
\end{eqnarray}

\begin{figure}[t!]
\includegraphics[width=8.6cm]{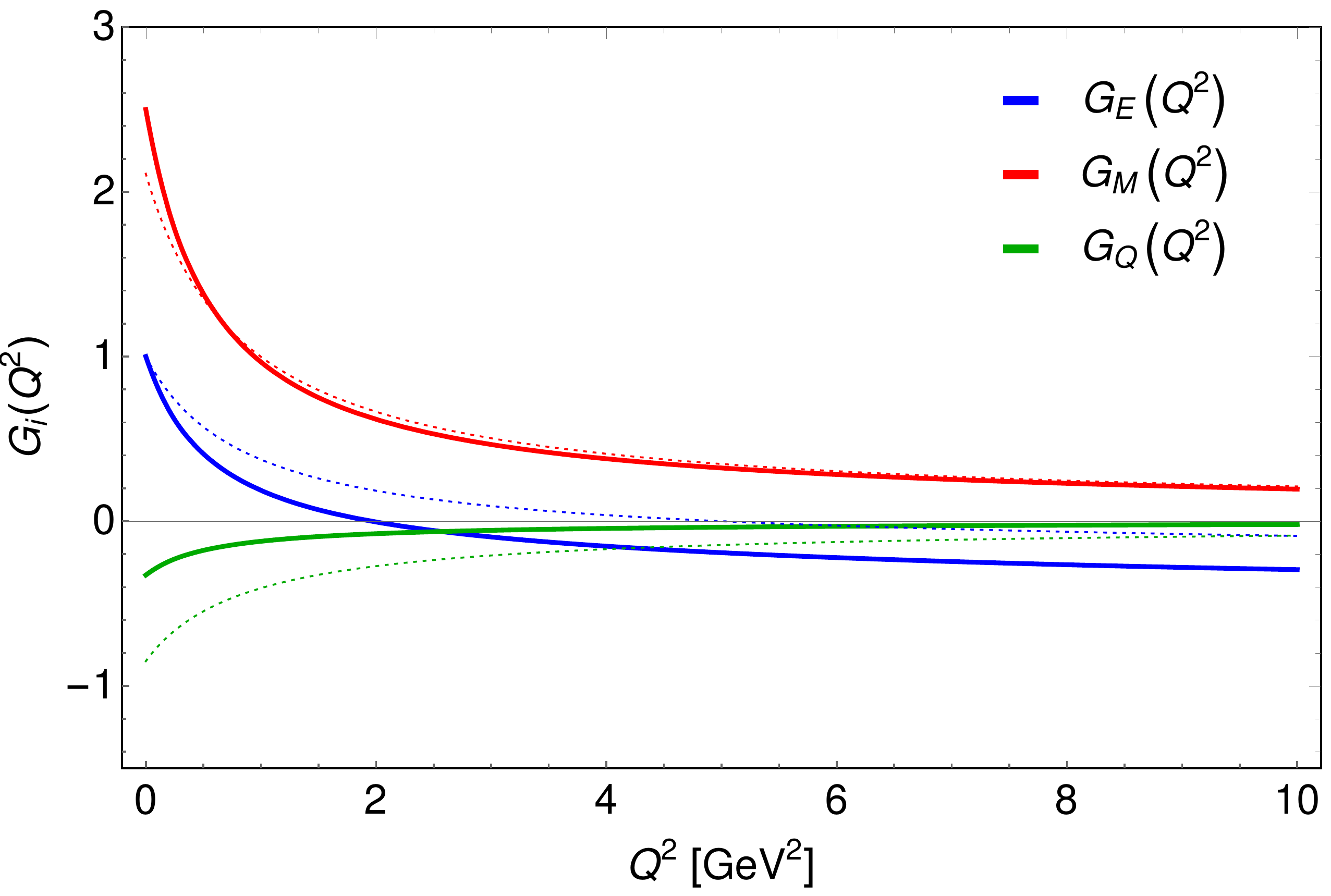}
\caption{$\rho$ \textit{meson form factors}. From top to bottom: $G_M$, $G_E$, $G_Q$. Solid thick lines are the results obtained in the CI-MRL truncation, whereas the dotted lines are their analogous in the CI-RL approach~\cite{Roberts:2011wy}.\label{fig:FFs}}
\end{figure}

\begin{figure}[t!]
\includegraphics[width=8.6cm]{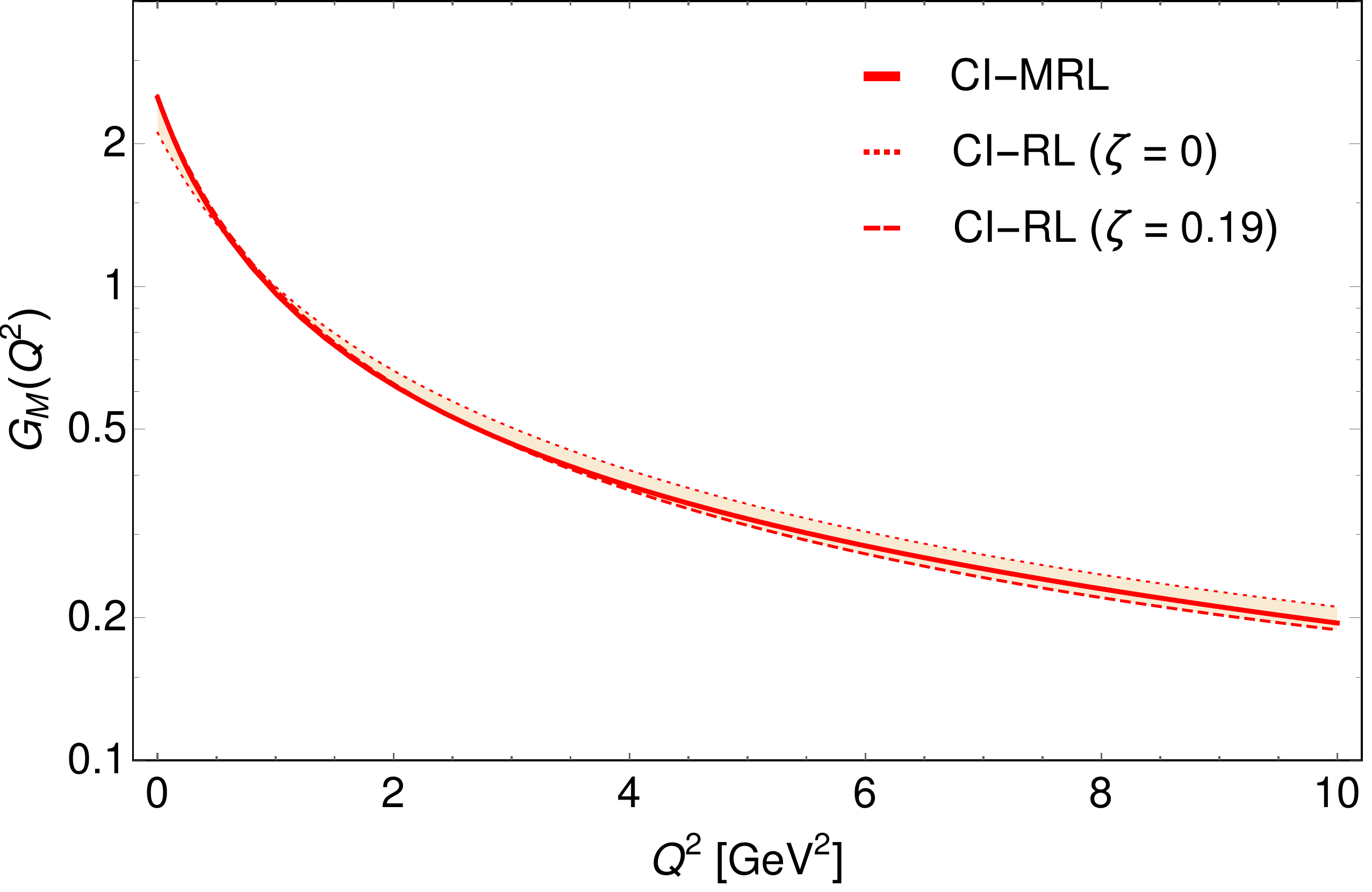}
\caption{$\rho$ \textit{meson magnetic form factor}. Solid thick line corresponds to the CI-MRL result. For the CI-RL case, we have employed the QPV Ansatz from Eqs.~(\ref{eq:AMMRL}, \ref{eq:EnhanceQPV}). With the strength parameter set to $\zeta_{\text{CI-RL}}=\zeta_{\text{CI-MRL}}=0.19$ in such case, the resemblance to the CI-MRL result is striking.  \label{fig:zoomGM}}
\end{figure}

The computed form factors are depicted in Fig.~\ref{fig:FFs}. The electric form factor $G_E$ exhibits a zero and remains negative thereafter. For the CI-MRL truncation, the zero is located at $Q^2 \approx 1.96$ GeV$^2$, while the CI-RL exhibits a zero at $Q^2 \approx 5$ GeV$^2$. This zero crossing was also observed in the DSE-BS approach from Ref.~\cite{Xu:2019ilh}. The magnetic form factor $G_M$, which turns out to be positive definite and monotonically decreasing, is quite similar in both truncations on the domain shown. The largest difference is  appreciated at low $Q^2$. This can be attributed to the lack of AMM term in the QPV for the CI-RL truncation, which enhances the value of $G_M$ in a vicinity of $Q^2 = 0$, without altering the large-$Q^2$ behavior. In Fig.~\ref{fig:zoomGM} we observe that, if the AMM piece is included in the CI-RL description, by using the vertex Ansatz from Eqs.~(\ref{eq:AMMRL}, \ref{eq:EnhanceQPV}), magnetic form factors become even more similar. The effects on $G_E$ and $G_Q$ are immaterial and not shown. The quadrupole form factor $G_Q$ is negative and decreases in magnitude as $Q^2$ increases; CI-MRL and CI-RL truncations produce results with notorious different magnitudes.

Some static properties that can be read from the form factors are collected in Table 2, namely charge, magnetic and quadrupole moments. For completeness, we have also included the corresponding radii, as defined from
\begin{equation}
    \textless r_i^2 \textgreater= -6 \frac{1}{G_i(0)} \frac{d G_i(Q^2)}{dQ^2}\Bigg|_{Q^2=0}\;.
\end{equation}

As can be inferred from Table 2, the MRL truncation produces results which are closer to those obtained from using sophisticated gluon models~\cite{Xu:2019ilh,Xu:2021mju} and lattice QCD~\cite{Owen:2015gva}, whereas the CI-RL gives results closer to the structureless meson limit. As explained before, the QPV Ansatz from Eqs.~(\ref{eq:AMMRL}, \ref{eq:EnhanceQPV}) makes the CI-RL magnetic form factor $G_M(Q^2)$ more similar to the CI-MRL case, and therefore the static properties derived from there.

\begin{table}[ht!]
\caption{\label{tab:static} $\rho$ \textit{meson static properties}, as obtained from its elastic electromagnetic form factors. The CI-RL results match those from Ref.~\cite{Roberts:2011wy}. In our current framework, it corresponds to the $\xi = 0$ limit of the CI-MRL truncation. For comparison, we have included novel DSE results from~\cite{Xu:2019ilh}. Additionally, Lattice QCD~\cite{Owen:2015gva} predicts $\mu_\rho = 2.21(8)$ and $r_E = 0.82(4)$ fm, and the structureless meson limit dictates $\mu_\rho = 2$ and $\mathcal{Q}_\rho = -1$.}
\begin{tabular}{c|ccc}
\hline
 &  \;CI-MRL\; & \;CI-RL\; & \;DSE\; \\
\hline
$r_E$/fm & 0.692 & 0.561 & 0.72 \\
$r_M$/fm & 0.603 & 0.515 & 0.69 \\
$r_{\mathcal{Q}}$/fm & 0.612 & 0.512 & - \\
\hline
$r_E m_\rho$ & 2.706 & 2.644 & 2.76 \\
$r_M m_\rho$ & 2.359 & 2.430 & 2.63 \\
$r_{\mathcal{Q}} m_\rho$ & 2.394 & 2.414 & - \\
\hline
$\mu_\rho$ & 2.500 & 2.110 & 2.01 \\
$\mathcal{Q}$ & -0.327 & -0.850 & -0.36 \\
\hline
\end{tabular}
\end{table}

\section{Distribution amplitudes}
To scrutinize a little further into the structure of the $\rho$ meson, and the effects of BS kernel truncation, we now examine the so called valence-quark distribution amplitudes. Particularly, let us consider two leading-twist distributions amplitudes of the $\rho$ meson: $\phi_\parallel(x)$ and $\phi_\perp(x)$. Intuitively, those describe the light-front momentum carried by the quark in a longitudinally or transversely polarised $\rho$,~\cite{Ball:1996tb}. In terms of quark propagators and BS amplitudes, the distributions can be written as~\cite{Gao:2014bca}:
\begin{eqnarray}
\nonumber
n\cdot P f_\rho  \phi_\parallel(x) && = m_{\rho} N_c \text{tr} \int_q \delta_n^x(q^+) \gamma \cdot n \; n_\nu \chi_\nu^\rho(q;P) \;, \\ \label{eq:PDAsdef}
f_\rho^{\perp} \phi_\perp(x) && = -\frac{1}{2} N_c \text{tr} \int_q \delta_n^x(q^+) n_{\mu}\sigma_{\mu\alpha} \mathcal{O}_{\alpha\nu}^{\perp} \chi_\nu^\rho(q;P)\;,
\end{eqnarray}
where $\delta_n^x(q^+):=\delta(n \cdot q^+ - x n \cdot P)$ and $\mathcal{O}_{\alpha\nu}^{\perp}=\delta_{\alpha\nu}+n_{\alpha}\bar{n}_{\nu}+\bar{n}_{\alpha}n_{\nu}$, and $n$ is a light-like four-vector such that $n^2 = 0$, $n\cdot P = -m_\rho$; $\bar{n}$ is a conjugate light-like four-vector, $\bar{n}^{2}=0$, $n\cdot \bar{n}=-1$; and, with the above definitions, $\int_0^1 dx \phi_\parallel(x) = \int_0^1 dx \phi_\perp(x) = 1$.
%

The PDAs can be obtained from their corresponding Mellin moments, $\textless x^m \textgreater = \int_0^1 dx x^m \phi(x)$.
After evaluating the Dirac trace, a series of algebraic manipulations involving Feynman parametrization, supplemented by the uniqueness of Mellin moments, enable us to identify
\begin{eqnarray}\nonumber
\phi_\parallel(x) &&=\frac{1}{f_{\rho}} \frac{N_c}{4\pi^2}m_\rho \Large[ 2x(1-x) E_\rho + F_\rho \Large]\;\bar{\mathcal{C}}_{1}^{\text{iu}}(\omega(M^2,x,-m_\rho^2))\;,\\
\phi_\perp(x) &&=\frac{1}{f_\rho^{\perp}} \frac{N_c}{4\pi^2}M  \left[ E_\rho + \left( 1+\frac{m_\rho^2}{M^2}x(1-x)\right)F_\rho \right] \nonumber \\
&& \times \;\bar{\mathcal{C}}_{1}^{\text{iu}}(\omega(M^2,x,-m_\rho^2))\;.\label{eq:PDAs}
\end{eqnarray}

\begin{figure}[t!]
\includegraphics[width=8.6cm]{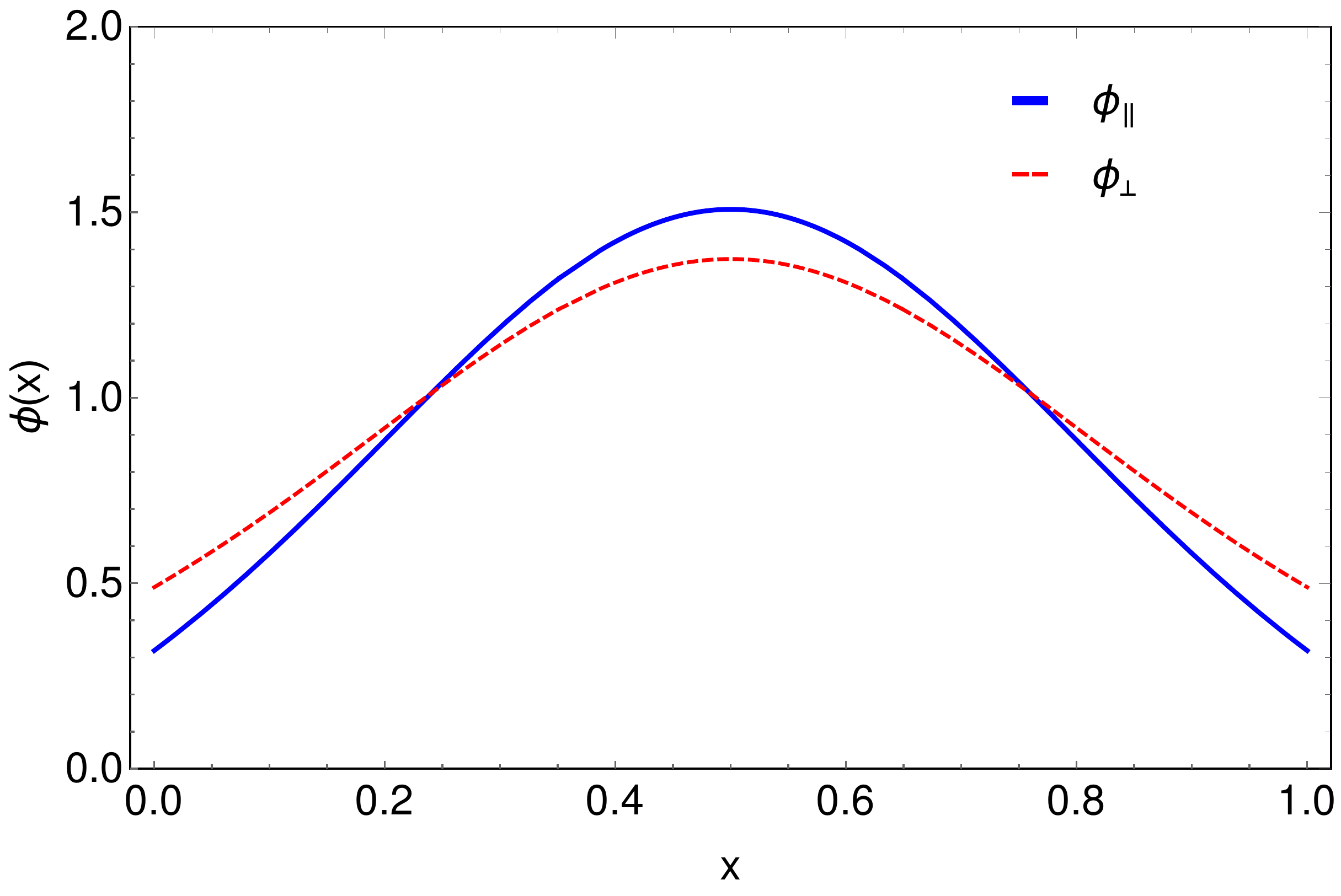}
\caption{$\rho$ \textit{meson PDAs}. Computed distributions, Eq.~\eqref{eq:PDAs}, in the CI-MRL truncation.\label{fig:PDA MRL}}
\end{figure}

\begin{figure}[t!]
\includegraphics[width=8.6cm]{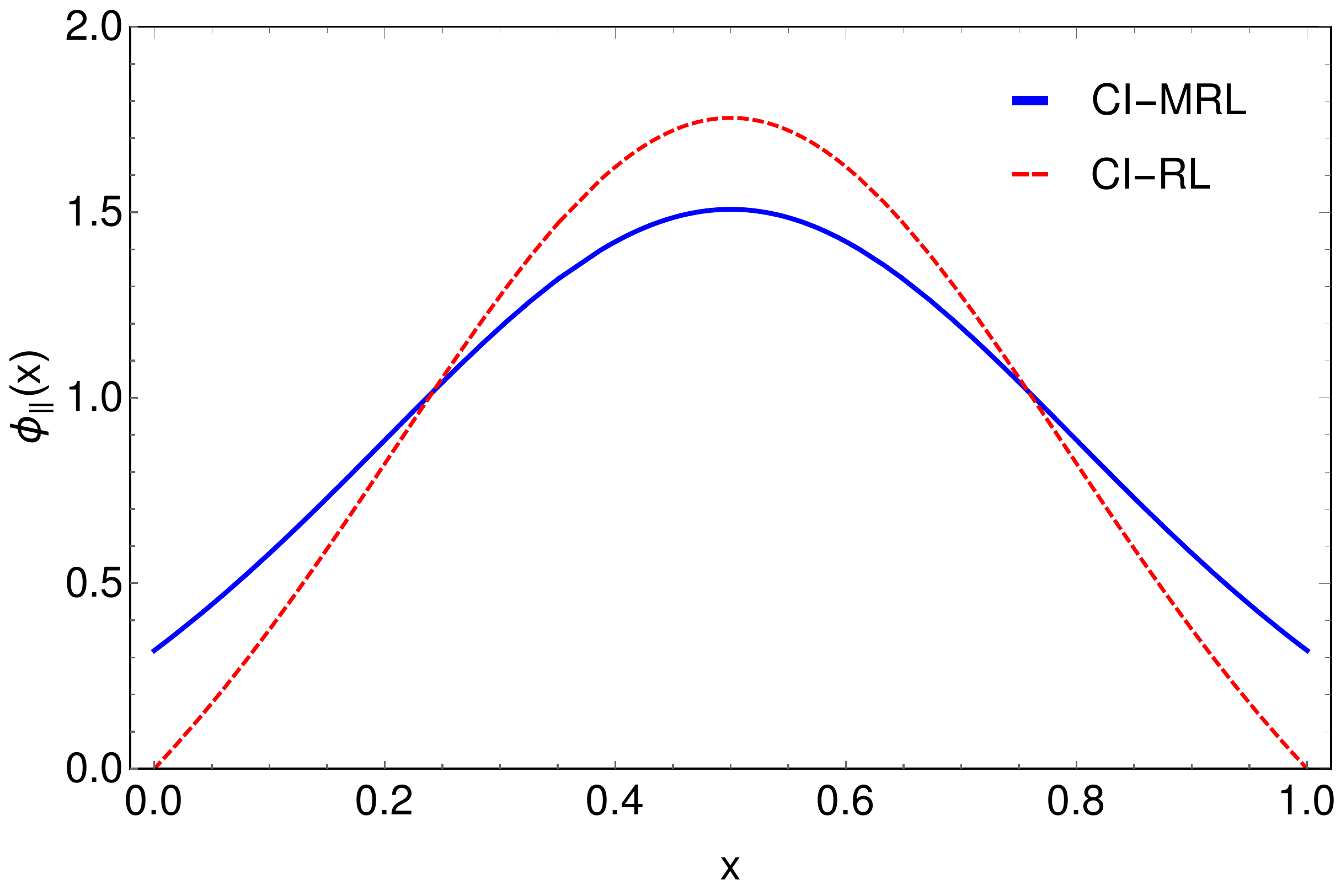}\\
\includegraphics[width=8.6cm]{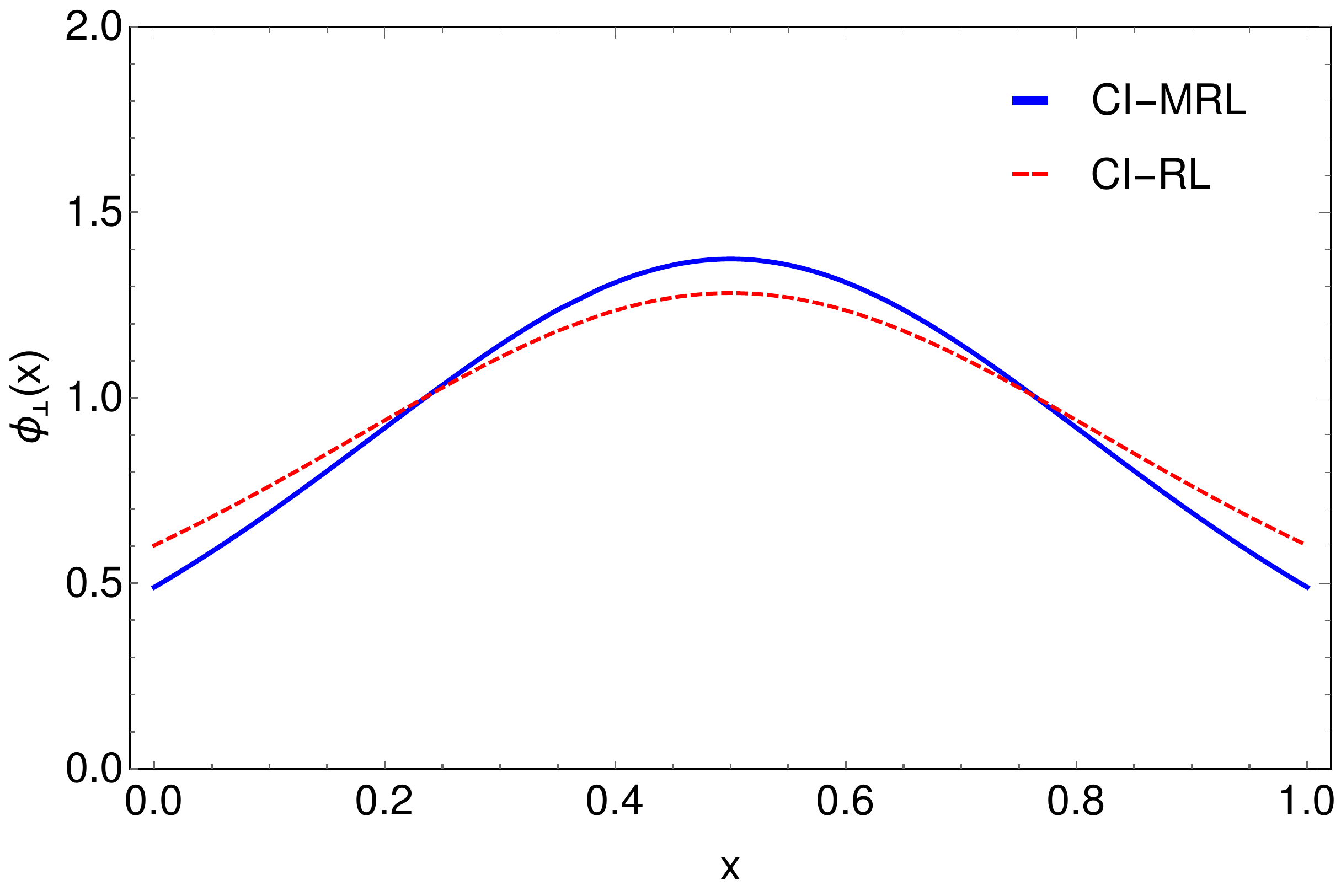}
\caption{$\rho$ \textit{meson PDAs}. Computed distributions, Eq.~\eqref{eq:PDAs}, in the CI-RL and CI-MRL truncations. Notably, $\phi_\parallel(x)$ in the CI-RL exhibits soft endpoint behavior.\label{fig:PDA comp}}
\end{figure}

The obtained PDAs are displayed in Fig.~\ref{fig:PDA MRL}. Both distributions manifest symmetry under the exchange $1 \leftrightarrow 1-x$, which is expected in the isospin symmetric limit $m_u = m_d$. The parallel distribution, $\phi_\parallel$, is more compressed with respect to its perpendicular counterpart, an expect pattern from sophisticated momentum-dependent interactions~\cite{Gao:2014bca}. It is evident that the derived PDAs do not vanish at the endpoints, in contradiction with QCD prescriptions~\cite{Gao:2014bca,Ball:1996tb,Lepage:1980fj}. The same occurs for the pion leading-twist PDA in the CI model, for which $\phi_\pi(x) = 1$ in the chiral limit~\cite{Roberts:2010rn}. Nonetheless, it is quite interesting that, in the absence of the $F_\rho$ BS amplitude (the CI-RL case), the distribution $\phi_\parallel$ actually exhibits soft endpoint behavior. This is well understood from Eqs.~\eqref{eq:PDAs} and drawn in Fig.~\ref{fig:PDA comp}.

\section{Conclusions}
Starting from the leading-order symmetry-preserving RL truncation for the two-body problem, we have illustrated how, although the vector and axial vector WGTIs provide necessary consistency conditions to construct the interrelated one body and two body kernels, these turn out insufficient to determine the kernels in an unambigous manner~\cite{Chang:2020iut}. Particularly, we construct a two-body kernel completely consistent with the aforementioned WGTIs that includes NL terms. The MRL truncation lets unaltered the pseudoscalar and axial vector meson properties, but yields some implications in the vector channel. This can be well illustrated by employing the CI model described in Section III. Among other things, we have seen how solutions of the QPV BS equations using the MRL kernel, automatically generate an anomalous magnetic moment term in the vertex structure. This is in addition to the vector meson pole at $Q^2=-m_\rho^2$, which is an artifact of the RL truncation~\cite{MiramontesLopez:2021oul}.  In the presence of DCSB, a dressed light-quark possesses a large anomalous electromagnetic moment~\cite{Chang:2010hb,Bashir:2011dp}. Therefore, the automatic incorporation of such term is an important outcome of the constructed two-body kernel in MRL truncation. Immediate consequences are naturally noted in the $\rho$ meson properties. For instance, although the produced decay constants are practically invariant under the effects of the NL pieces we have introduced in the BS kernel, the experimental mass of the $\rho$ can be faithfully obtained within the CI-MRL, in contrast with the CI-RL approach which produces a larger value~\cite{Roberts:2011wy}. In some way, the NL term of the two-body kernel mimicks the effects of the pion cloud effects, which correct the inflated masses produced when they are not considered~\cite{Eichmann:2008ef};
whether or not our kernel is actually related to pion cloud effects is currently under investigation. On another note, whereas the CI-RL truncation yields static properties (as derived from the corresponding EFF) closer to the structureless meson limit, CI-MRL results are more compatible with those obtained from more sophisticated approaches~\cite{Xu:2019ilh,Owen:2015gva}. Focusing on the magnetic form factor, we have reinforced the importance of the AMM for a good description of vector mesons. Finally, we have derived the $\rho$ meson $\phi_\parallel(x)$ and $\phi_\perp(x)$ distribution amplitudes. The necessity of employing a regularization scheme, and subsequently cut-offs, to deal with the CI model, makes the PDAs not vanishing at the endpoints. This is also observed in the pion case: the chiral limit pseudoscalar yields $\phi_\pi(x) =1$~\cite{Roberts:2010rn}. Interestingly, in the absence of the $F_\rho$ piece of the Bethe-Salpeter amplitude (the CI-RL case), the distribution $\phi_\parallel(x)$ actually vanishes at the endpoints. Although it is an expected behavior from QCD grounds,~\cite{Gao:2014bca}, it seems rather incompatible with the pion result and the CI framework in general. Having exposed some impacts of the two-body kernel on the QPV structure and $\rho$ meson properties, via masses and decay constants, electromagnetic form factors and distribution amplitudes, the question on its implications on the distribution functions remains unanswered. This is an aspect of the $\rho$ meson structural properties that will be addressed elsewhere. A similar study for the open flavor case, in particular the Kaon sector, will be conducted as well.

\begin{acknowledgments}
The authors acknowledge valuable comments from José Rodríguez-Quintero. 
\end{acknowledgments}
\bibliography{main}

\appendix\widetext
\section{\label{App:FFs}Appendix $\rho$ form factors}
The vertex characterizing the elastic EFFs of the $\rho$ meson is written in Eq.~\eqref{eq:lambdavertex2}. With properly chosen projection operators, the three form factors $F_i(Q^2)$ are written, in general:
\begin{eqnarray}
F_i=\int_{0}^{1}du_1\int_{0}^{1-u1}du_2 \mathcal{A}_i\bar{\mathcal{C}}_{1}^{\text{iu}}(\Omega)+\mathcal{B}_i\bar{\mathcal{C}}_{2}^{\text{iu}}(\Omega)
\end{eqnarray}
where $\Omega(u_1,u_2,M^2,P^2,Q^2)=M^2-P^2(u_1+u_2)(u_1+u_2-1)+u_1 u_2 Q^2$ and $\bar{\mathcal{C}}_{2}^{\text{iu}}(\Omega)=\frac{d^2}{d\Omega^2}\mathcal{C}^{\text{iu}}(\Omega)$. $\mathcal{X}_i=\mathcal{X}_i^{EE}(Q^2)E_c^2+\mathcal{X}_i^{EF}(Q^2)E_cF_c+\mathcal{X}_i^{FF}(Q^2)F_c^2$, for $\mathcal{X}=\mathcal{A}, \mathcal{B}$ and $i=1,2,3$. Explicitly, we arrive at the following expressions for the different pieces:
\begin{eqnarray}
&&\mathcal{A}_1^{EE}(Q^2)=-\frac{N_c}{4\pi^2}\left\{\frac{Q^2(u_1-u_2)(6m_\rho^4+5m_\rho^2Q^2+Q^4)}{4m_\rho^4(Q^2+4m_\rho^2)}V_1(Q^2)+\left(u_1+u_2-2\right)V_2(Q^2)\right\}\nonumber\\
&&\mathcal{A}_1^{EF}(Q^2)=\frac{N_c}{4\pi^2}\left\{4V_2(Q^2)-\frac{Q^2(u_1+u_2)}{M^2}V_3(Q^2)\right\}\nonumber\\
&&\mathcal{A}_1^{FF}(Q^2)=\frac{N_c}{4\pi^2}\left\{\frac{Q^2(u_1-u_2)(Q^2+2m_\rho^2)}{4M^2(Q^2+4m_\rho^2)}V_1(Q^2)+\frac{m_\rho^2}{M^2}\left(u_1+u_2\right)V_2(Q^2)\right\}
\end{eqnarray}

\begin{eqnarray}
&&\mathcal{A}_2^{EE}(Q^2)=-\frac{N_c}{4\pi^2}\left\{\frac{2(u_1-u_2)(Q^2+3m_\rho^2)}{Q^2+4m_\rho^2}V_1(Q^2)+2\left(u_1+1\right)V_2(Q^2)+4V_3(Q^2)\right\}\nonumber\\
&&\mathcal{A}_2^{EF}(Q^2)=\frac{N_c}{4\pi^2}\left\{-4V_2(Q^2)+\frac{4m_\rho^2(u_1+u_2-1)+2u_2Q^2}{M^2}V_3(Q^2)\right\}\nonumber\\
&&\mathcal{A}_2^{FF}(Q^2)=\frac{N_c}{4\pi^2}\left\{\frac{2m_\rho^4(u_1-u_2)}{M^2(Q^2+4m_\rho^2)}V_1(Q^2)-\frac{2u_2m_\rho^2}{M^2}V_2(Q^2)\right\}
\end{eqnarray}

\begin{eqnarray}
&&\mathcal{A}_3^{EE}(Q^2)=\frac{N_c}{4\pi^2}\left\{\frac{(u_1-u_2)(60m_\rho^6+32m_\rho^4Q^2+7m_\rho^2Q^4+Q^6)}{m_\rho^2(Q^2+4m_\rho^2)^2}V_1(Q^2)+4\frac{m_\rho^2(u_1-u_2)}{Q^2+4m_\rho^2}V_2(Q^2)\right\}\nonumber\\
&&\mathcal{A}_3^{EF}(Q^2)=\frac{N_c}{4\pi^2}\left\{\frac{4m_\rho^2Q^2(u_1-u_2)}{M^2(Q^2+4m_\rho^2)}V_3(Q^2)\right\}\nonumber\\
&&\mathcal{A}_3^{FF}(Q^2)=\frac{N_c}{4\pi^2}\left\{-\frac{m_\rho^2(u_1-u_2)(20m_\rho^4+4m_\rho^2Q^2+Q^4)}{M^2(Q^2+4m_\rho^2)^2}V_1(Q^2)+\frac{4m_\rho^4(u_2-u_1)}{M^2(Q^2+4m_\rho^2)}V_2(Q^2)\right\}
\end{eqnarray}

\begin{eqnarray}
&&\mathcal{B}_1^{EE}(Q^2)\nonumber\\
=&&\frac{N_c}{4\pi^2}\left\{\frac{Q^2(u_1-u_2)(Q^2+2m_\rho^2)}{8m_\rho^4(Q^2+4m_\rho^2)}\right.\nonumber\\
\times&&\left.\left[-M^2(Q^2+3m_\rho^2)+m_\rho^4(u_1+u_2-1)(9u_1+9u_2-2)+m_\rho^2Q^2((7u_1-3)u_2+3(u_1-1)u_1+3u_2^2)+u_1u_2Q^4\right]V_1(Q^2)\right.\nonumber\\
+&&\left.\frac{1}{2}\left[M^2(2-u_1-u_2)+m_\rho^2(u_1+u_2-1)(u_1+u_2)(3u_1+3u_2-4)+3u_1u_2(u_1+u_2-2)Q^2\right]V_2(Q^2)-Q^2V_3(Q^2)\right\}\nonumber\\
\nonumber\\
&&\mathcal{B}_1^{EF}(Q^2)\nonumber\\
=&&\frac{N_c}{4\pi^2}\left\{-\frac{Q^2(u_1-u_2)(Q^2+2m_\rho^2)}{4m_\rho^2}V_1(Q^2)-\frac{8m_\rho^4(u_1+u_2-1)(2u_1+2u_2+1)+4m_\rho^2Q^2(4u_1u_2+u_1+u_2)}{4m_\rho^2}V_2(Q^2)\right.\nonumber\\
+&&\left.\frac{Q^2((u_1+u_2)(m_\rho^2(u_1+u_2-1)(3u_1+3u_2+2)+3u_1u_2Q^2)-M^2(u_1+u_2+4))}{2M^2}V_3(Q^2)\right\}\nonumber\\
&&\mathcal{B}_1^{FF}(Q^2)\nonumber\\
\nonumber\\
=&&\frac{N_c}{4\pi^2}\left\{\frac{Q^2(u_1-u_2)(Q^2+2m_\rho^2)(-3M^2+m_\rho^2(u_1+u_2-1)(u_1+u_2+2)+u_1u_2Q^2)}{8M^2(Q^2+4m_\rho^2)}V_1(Q^2)\right.\nonumber\\
+&&\left.\left[\frac{m_\rho^2(m_\rho^2(u_1+u_2-2)(u_1+u_2-1)(u_1+u_2)+Q^2(-u_1^2u_2+2u_1^2(u_1-1)-(2+u_1)u_2^2+2u_2^3))}{2M^2}\right.\right.\nonumber\\
-&&\left.\left.\frac{1}{2}(m_\rho^2(3u_1+3u_2-4)+2Q^2(u_1+u_2))\right]V_2(Q^2)\right.\nonumber\\
+&&\left.\frac{Q^2(-M^2+m_\rho^2(u_1+u_2-2)(u_1+u_2)+u_1u_2Q^2)}{M^2}V_3(Q^2)\right\}
\end{eqnarray}

\begin{eqnarray}
&&\mathcal{B}_2^{EE}(Q^2)\nonumber\\
=&&\frac{N_c}{4\pi^2}\left\{\frac{(u_1-u_2)}{Q^2+4m_\rho^2}\right.\nonumber\\
\times&&\left.\left[-M^2(Q^2+3m_\rho^2)+m_\rho^4(u_1+u_2-1)(9u_1+9u_2-2)+m_\rho^2Q^2((7u_1-3)u_2+3(u_1-1)u_1+3u_2^2)+u_1u_2Q^4\right]V_1(Q^2)\right.\nonumber\\
+&&\left.\left[-(u_1+1)M^2+m_\rho^2(u_1+u_2-1)(3u_1^2+3u_1u_2+u_1+3u_2)+u_1(3u_1+1)u_2Q^2\right]V_2(Q^2)\right.\nonumber\\
+&&\left.\left[2m_\rho^2(u_1+u_2-1)(2u_1+2u_2+1)+2u_1(2u_2+1)Q^2\right]V_3(Q^2)\right\}\nonumber\\
\nonumber\\
&&\mathcal{B}_2^{EF}(Q^2)\nonumber\\
=&&\frac{N_c}{4\pi^2}\left\{\left[2m_\rho^2(u_2-u_1)\right]V_1(Q^2)+\left[2(m_\rho^2(2u_1^2+u_1(4u_2-3)+u_2(2u_2-3)-1)+2u_1u_2Q^2)\right]V_2(Q^2)\right.\nonumber\\
+&&\left.\left[\frac{-6m_\rho^4(u_1+u_2-1)^2(u_1+u_2)+m_\rho^2Q^2(-2u_1^3+u_1^2(2-5u_2)+u_1(7-6u_2)u_2-3(u_2-1)u_2^2)-u_1u_2Q^4}{M^2}\right.\right.\nonumber\\
+&&\left.\left.2m_\rho^2(u_1+u_2-3)+Q^2(2u_1+u_2)\right]V_3(Q^2)\right\}\nonumber\\
\nonumber\\
&&\mathcal{B}_2^{FF}(Q^2)\nonumber\\
=&&\frac{N_c}{4\pi^2}\left\{\frac{m_\rho^4(u_1-u_2)(-3M^2+m_\rho^2(u_1+u_2-1)(u_1+u_2+2)+u_1u_2Q^2)}{M^2(Q^2+4m_\rho^2)}V_1(Q^2)\right.\nonumber\\
+&&\left.\left[\frac{m_\rho^2(-(u_2+2)M^2+m_\rho^2(u_1+u_2-1)(3u_1u_2+2u_1+3u_2^2)+3u_1u_2^2Q^2)}{M^2}\right]V_2(Q^2)\right.\nonumber\\
+&&\left.\frac{2m_\rho^2(-M^2+m_\rho^2(u_1+u_2-1)(u_1+u_2+1)+(u_1+1)u_2Q^2)}{M^2}V_3(Q^2)\right\}
\end{eqnarray}

\begin{eqnarray}
&&\mathcal{B}_3^{EE}(Q^2)\nonumber\\
=&&\frac{N_c}{4\pi^2}\left\{\frac{-(u_1-u_2)(20m_\rho^4+4m_\rho^2Q^2+Q^4)}{2m_\rho^2(Q^2+4m_\rho^2)^2}\right.\nonumber\\
\times&&\left.\left[-M^2(Q^2+3m_\rho^2)+m_\rho^4(u_1+u_2-1)(9u_1+9u_2-2)+m_\rho^2Q^2((7u_1-3)u_2+3(u_1-1)u_1+3u_2^2)+u_1u_2Q^4\right]V_1(Q^2)\right.\nonumber\\
+&&\left.\frac{2m_\rho^2(M^2(u_1-u_2)-m_\rho^2(u_1+u_2-1)(2(8u_1+1)u_2+u_1(3u_1-2)-3u_2^2)-u_1u_2Q^2(7u_1+u_2-4))}{Q^2+4m_\rho^2}V_2(Q^2)\right.\nonumber\\
-&&\left.\frac{4m_\rho^2(4m_\rho^2(u_1+u_2-1)+(2u_1-1)Q^2)}{Q^2+4m_\rho^2}V_3(Q^2)\right\}\nonumber\\
\nonumber\\
&&\mathcal{B}_3^{EF}(Q^2)\nonumber\\
=&&\frac{N_c}{4\pi^2}\left\{\frac{(u_1-u_2)(20m_\rho^4+4m_\rho^2Q^2+Q^4)}{Q^2+4m_\rho^2}V_1(Q^2)+\left[4m_\rho^2(u_1+u_2)\right]V_2(Q^2)\right.\nonumber\\
+&&\left.\frac{2m_\rho^2}{M^2(Q^2+4m_\rho^2)}\left[-M^2(8m_\rho^2(u_1+u_2-2)+Q^2(3u_1+u_2-4))+8m_\rho^4(u_1+u_2-1)^2(u_1+u_2)\right.\right.\nonumber\\
+&&\left.\left.m_\rho^2Q^2(u_1+u_2)(-4u_1u_2+(u_1-3)u_1+3u_2^2-5u_2+2)-u_1u_2Q^4(3u_1+u_2)\right]V_3(Q^2)\right\}\nonumber\\
\nonumber\\
&&\mathcal{B}_3^{FF}(Q^2)\nonumber\\
=&&\frac{N_c}{4\pi^2}\left\{-\frac{m_\rho^2(u_1-u_2)(20m_\rho^4+4m_\rho^2Q^2+Q^4)(-3M^2+m_\rho^2(u_1+u_2-1)(u_1+u_2+2)+u_1u_2Q^2)}{2M^2(Q^2+4m_\rho^2)^2}V_1(Q^2)\right.\nonumber\\
+&&\left.\frac{2m_\rho^2}{M^2(Q^2+4m_\rho^2)}\left[M^2(m_\rho^2(7u_1+9u_2)+2Q^2(u_1+u_2))+m_\rho^2(Q^2(-2u_1^3+u_1^2(u_2+2)-5u_1u_2^2-2(u_2-1)u_2^2)\right.\right.\nonumber\\
-&&\left.\left.m_\rho^2(u_1+u_2-1)(u_1(5u_1+2)+u_2(11u_2-2)))\right]V_2(Q^2)\right.\nonumber\\
-&&\left.\frac{4m_\rho^2(-M^2(Q^2+4m_\rho^2)+4m_\rho^4(u_1+u_2-1)(u_1+u_2)+m_\rho^2Q^2(6u_1u_2+(u_1-2)u_1+u_2^2)+u_1u_2Q^4)}{M^2(Q^2+4m_\rho^2)}V_3(Q^2)\right\}
\end{eqnarray}
\end{document}